\documentclass[aps,prb,superscriptaddress, twocolumn]{revtex4}
\usepackage[dvips]{graphicx}
\usepackage{setspace} 

\begin{document}

\title{The change of electronic state and crystal structure by post-annealing in superconducting SrFe$_2$(As$_{0.65}$P$_{0.35}$)$_2$}

\author{T.~Kobayashi}
\affiliation{Department of Physics, Osaka University, Osaka 560-0043, Japan.}

\author{S.~Miyasaka}
\affiliation{Department of Physics, Osaka University, Osaka 560-0043, Japan.}

\author{S.~Tajima}
\affiliation{Department of Physics, Osaka University, Osaka 560-0043, Japan.}

\author{T.~Nakano}
\affiliation{Department of Physics, Osaka University, Osaka 560-0043, Japan.}

\author{Y.~Nozue}
\affiliation{Department of Physics, Osaka University, Osaka 560-0043, Japan.}

\author{N.~Chikumoto}
\affiliation{Supercunductivity Research Laboratory-ISTEC, Tokyo 135-0062, Japan}

\author{H.~Nakao}
\affiliation{Condenced Matter Research Center and Photon Factory, Institute of Materials Structure Science, High Energy Accelerator Research Organization, Tsukuba 305-0801, Japan}

\author{R.~Kumai}
\affiliation{Condenced Matter Research Center and Photon Factory, Institute of Materials Structure Science, High Energy Accelerator Research Organization, Tsukuba 305-0801, Japan}

\author{Y.~Murakami}
\affiliation{Condenced Matter Research Center and Photon Factory, Institute of Materials Structure Science, High Energy Accelerator Research Organization, Tsukuba 305-0801, Japan}

\setcounter{tocdepth}{3}
%%%%%%%%%%%%%%%%%%%%%%% abstract %%%%%%%%%%%%%%%%%%%%%%%%%%%%%%%%%%
\begin{abstract}
We investigated the annealing effects on the physical properties of SrFe$_2$(As$_{1-x}$P$_x$)$_2$ ($x=0.35$). The superconducting transition temperature ($T_c$) increased from 26~K to 33~K by annealing. The X-ray diffraction measurement suggested that the annealed crystals have the shorter/longer $a/c$-axes and the larger pnictogen height $h_\mathrm{Pn}$. This must be linked to the $T_c$-enhancement by annealing. Moreover, it was found that the post-annealing decreased the electronic specific heat coefficient at $T$=0~K, $\gamma_r$, and changed the magnetic field ($H$) dependence from sub-linear $\gamma_r\propto H^{0.7}$ to $H$-linear $\gamma_r\propto H$. This can be attributed the electronic change from dirty to clean superconductors with $s_{\pm}$ gap.
\end{abstract}

%%%%%%%%%%%%%%%%%%%%%%%%%%%%%%%%%
\maketitle
%%%%%%%%%%%%%%%%%%%%%%%%%%%%%%%%%%
%\draft
%\linenumbers
%\narrowtext
%%%%%%%%%%%%%%%%%%%%%introduction%%%%%%%%%%%%%%%%%%%%%%%%%%%%%
%\section{Introduction}
~~In order to determine the pairing interaction in iron based superconductors (FeSC), it is important to clarify the superconducting gap structure. Intensive researches so far have revealed that most of the FeSC are full gap superconductors, while some materials, such as LaFePO, KFe$_2$As$_2$, and P-doped BaFe$_2$As$_2$ (P-Ba122), are nodal superconductors\cite{D. C. Johnston, G. R. Stewart}. For example, in P-Ba122 the nodal gap behavior was observed by NMR\cite{Y. Nakai}, penetration depth and thermal conductivity measurement\cite{K. Hashimoto}, while the topology and the location of the node is still under debates \cite{M. Yamashita, Y. Zhang}.

~~Specific heat ($C$) measurement is sensitive to a low energy excitation, and thus a good probe for a superconducting gap structure. In a full gap superconductor, the residual part of electronic specific heat coefficient $\gamma_r$($\equiv$ $C/T$ at zero temperature) under magnetic field $H$ is proportional to $H$, while $\gamma_r\propto H^{0.5}$ was predicted in a nodal superconductor by Volovik\cite{G. E. Volovik} and was confirmed in cuprate superconductors\cite{K. A. Moler}. However, in multiband systems, such as FeSC, the magnetic field dependence of $\gamma_r(H)$ also depends on the gap ratio. Moreover, in a nodal or $s_{\pm}$ gap superconductor $\gamma_r(H)$ is sensitive to the cleanness of the system. 
Bang pointed out that both of $\gamma_r\propto H$ and $\gamma_r\propto H^{0.5}$ can be explained by the $s_{\pm}$ full gap scheme with impurities\cite{Y. Bang}. Therefore, high quality crystals are required to clarify the intrinsic nature of superconducting gap. 
On the other hand, it was found that post-annealing of as-grown crystals sometimes gives a remarkable change in the electronic properties of FeSC. For example, the large anisotropy in low-$T$ in-plane resistivity of BaFe$_2$As$_2$ disappeared after annealing\cite{S. Ishida}. 
It suggests some intrinsic change in the electronic state with annealing treatment. Recently we found that $T_c$ is substantially enhanced by post-annealing in SrFe$_2$(As, P)$_2$ (P-Sr122), while the annealing effects are not remarkable in P-Ba122\cite{unpublished}. 
Since the electronic properties including superconductivity are quite sensitive to a small structural change in FeSC, we need to study carefully the annealing effects on the electronic properties.
 
~~Here, we focus on P-Sr122 to clarify the annealing effect on the crystal and electronic structure, including the gap nature. 
In the previous study, we reported that the crystal structure of P-Sr122 is more three dimensional than that of P-Ba122, and resistivity showed $T$-linear behavior indicating two dimensional antiferromagnetic (2D-AFM) fluctuation \cite{T. Kobayashi}. 
The NMR measurement on as-grown crystals found that $1/T_{1}T = const.$ below $T_c$, indicating a line node of the gap\cite{T. Dulguun}. 
The penetration depth measurement showed $\Delta\lambda(T)\propto T^n~(n=1.5\sim2)$, whose exponent $n$ is also consistent with a nodal gap\cite{J. Murphy, H. Takahashi}.

~~In this study, we compare magnetic susceptibility, electric resistivity, specific heat and X-ray diffraction analysis for as-grown and annealed SrFe$_2$(As$_{0.65}$P$_{0.35}$)$_2$ single crystals. 
We found that the annealing effect is  very strong in P-Sr122 system. The $T_c$ of the as-grown crystals is $T_c$=26~K, while the annealing treatment raises it up to 33~K that is higher than the optimal value for P-Ba122 ($T_c$=31~K) \cite{S. Kasahara}. 
The $\gamma_r (H)$ of the as-grown crystal showed sub-linear $H$-dependence, while the annealed crystal showed $H$-linear dependence with a smaller residual term. 
In addition, the average crystal structure seems to change with post-annealing.
%%%%%%%%%%%%%%%%%%%%%%%%%%experiment detail%%%%%%%%%%%%%%%%%%%%%%%%%%%
%\section{Experimental Details}

~~Single crystals of SrFe$_2$(As$_{1-x}$P$_x$)$_2$ ($x=0.35$) were grown from a stoichiometric mixture of Sr, FeAs, FeP powder in an alumina crucible, sealed in a silica tube with Ar gas of 0.2~bar at room temperature. It was heated up to 1300~$^\circ$C, kept for 12~hours, and then slowly cooled down to 1050~$^\circ$C at a rate of 2~$^\circ$C/h. To find an optimal
annealing condition, the as-grown crystals were annealed at various temperatures (400$\sim$800~$^\circ$C) for various times (1$\sim$14~days). We found that the annealing treatment at 500~$^\circ$C gave the highest $T_c$ among all our examinations, while the optimum annealing time depends on the crystal size.  For the specific heat measurement the crystals with a size of $3\times 3\times 0.1~$mm$^3$ were annealed in an evacuated silica tube for 2~weeks at 500~$^\circ$C, while the crystals for the other measurements were $1\times 1\times 0.1~$mm$^3$ in size and annealed for 1 week. It is noted that such an annealing temperature and time dependence was previously reported\cite{J. S. Kim2}. The electric resistivity was measured by a standard four probe method and the magnetic susceptibility was measured by SQUID. Specific heat was measured down to 1.8~K in magnetic field up to 14~T using a Quantum Design Physical Properties Measurement System (PPMS). X-ray diffraction experiment for single crystals was carried out using the X-ray with 15~keV at BL-8A of the Photon Factory, KEK in Japan. We used the atomic positions of SrFe$_2$As$_2$ as starting parameters\cite{S. R. Saha} and refined them by the least squares method using Rigaku CrystalStructure. 
                   
%%%%%%%%%%%%%%%%%experiment result and discussion%%%%%%%%%%%%%%%%%%%%%
%\section{Experiment result and Discussion}
\begin{figure}[h]
\begin{center}
\hspace{2mm}\includegraphics[width=7cm]{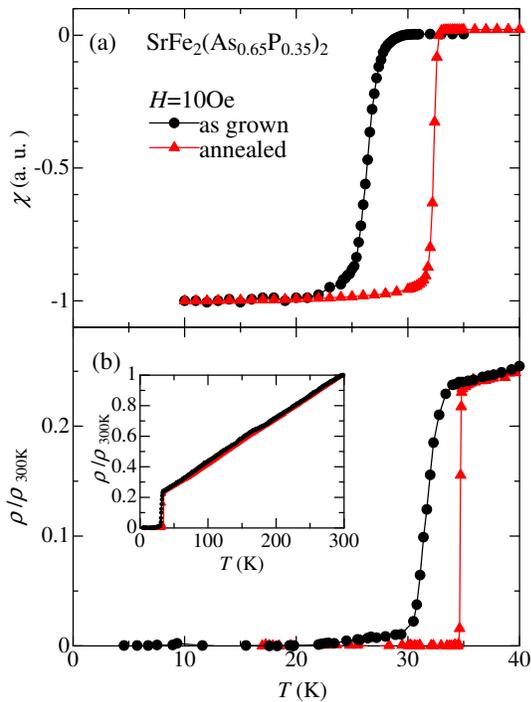}\\
%\vspace{1mm}
%\includegraphics[width=8.5cm]{eps/Fig_01_test2.eps}
\caption{
(Color online)(a) Temperature dependence of magnetic susceptibility $\chi$ for the as-grown (closed black circles) and the annealed (closed red triangles) SrFe$_2$(As$_{0.65}$P$_{0.35}$)$_2$, respectively. The data are normalized at 10~K. (b) Temperature dependence of resistivity normalized by the room temperature value. The inset shows resistivity over a wider $T$-range. The solid lines are guides for the eyes.
}
\label{fig1}
\end{center}
\end{figure}
%%%%%%%%%%%%%%%%resistivity and magnetization %%%%%%%%%%%%%%%%%%%%%%%%%%%%%%%%%

~~Figures 1(a) and (b) present the temperature dependences of magnetic susceptibility and electric resistivity, respectively. We defined $T_c$ at the middle of superconducting transition in susceptibility $\chi(T)$. In Figs. 1(a) and (b), the annealed sample shows a sharper transition at the higher $T_c$=33~K than that of the as-grown crystal with $T_c$=26~K. As shown in the inset of Fig. 1(b), both of the as-grown and annealed crystals show a $T$-linear resistivity in a wide temperature range, suggesting a strong 2D-AFM fluctuation. Such a 2D-AFM fluctuation was also observed by the NMR measurement\cite{T. Dulguun}. Residual resistivity ratio (RRR) slightly increased from 7.06 to 7.24 by annealing, which indicates a suppression of carrier scattering by disorders. The absolute value of resistivity at room temperature is 0.22~$\pm$~0.03~(m$\Omega$\, cm) and does not change by annealing within a measurement error.

%%%%%%%%%%%%%%%%%%%%%%%%%%%%%%%%%%%%%%%%%%%%%%%%%%%%%%%%%%%%%%

%%%%%%%%%%%%%%%%specific heat%%%%%%%%%%%%%%%%%%%%%%%%%%%%%%%%%
\begin{figure}[h]
\begin{center}
\hspace{2mm}\includegraphics[width=7cm]{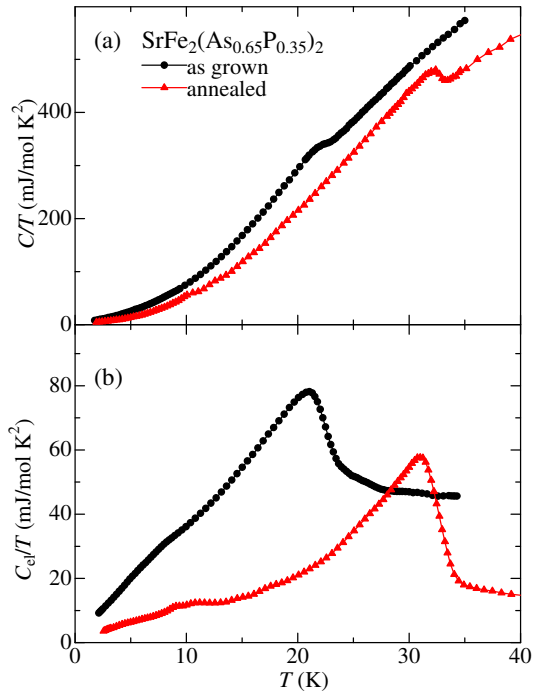}\\
%\vspace{1mm}
%\includegraphics[width=8.5cm]{eps/Fig_01_test2.eps}
\caption{
(Color online)(a) The temperature dependence of $C/T$ for the as-grown (black circles) and the annealed (red triangles) SrFe$_2$(As$_{0.65}$P$_{0.35}$)$_2$. (b) The temperature dependence of $C_{el}/T$ for as-grown and annealed crystals.  
}
\label{fig2}
\end{center}
\end{figure}
%%%%%%%%%%%%%%%%%%%%%%%%%%%%%%%%%%%%%%%%%%%%%%%%%%%%%%%%%%%%%%%
~~In Fig. 2(a), we plotted the temperature dependence of specific heat $C(T)$ divided by $T$. In order to subtract a phonon contribution, we used the phonon specific heat of non-superconducting SrFe$_2$As$_2$ and SrFe$_2$P$_2$. The phonon terms of specific heat for these end members ($C_\mathrm{phonon(SrFe_2As_2/SrFe_2P_2)}$) were estimated by the formula of 
$
C_\mathrm{phonon(SrFe_2As_2/SrFe_2P_2)}=C-\gamma_\mathrm{(SrFe_2As_2/SrFe_2P_2)}T.
$
Here, $C$ and $\gamma_\mathrm{(SrFe_2As_2/SrFe_2P_2)}$ are total specific heat and electronic specific heat coefficient for these compounds, respectively.
For the annealed crystal, $C_{phonon}$ was calculated as a combination of the values of SrFe$_2$As$_2$ and SrFe$_2$P$_2$, 
\[
C_{phonon}=0.7\times C_\mathrm{{phonon(SrFe_2As_2)}}+0.3\times C_\mathrm{{phonon(SrFe_2P_2)}},
\]
while $C_{phonon}$ was assumed to be equal to $C_{phonon(\rm SrFe_2As_2)}$ for the as-grown crystal. Figure 2(b) shows only the extracted electronic component of $C/T$. 
Here, $C_{el}$ for SrFe$_2$(As$_{0.65}$P$_{0.35})_2$ is estimated by the formula of $C_{el}=C-C_{\mathrm{phonon}}$. 
A jump due to superconducting transition was observed at 23~K for the as-grown crystal and a sharper jump at 33~K for the annealed crystal. 
The normal state Sommerfeld coefficient also changed from 47~mJ/mol~K$^2$ to 17~mJ/mol~K$^2$ by annealing. It seems that the post-annealing suppresses the mass enhancement. However, such a large change of the normal state Sommerfeld coefficient by the annealing treatment was not observed in the previous study of annealed Co-doped BaFe$_2$As$_2$ (Co-Ba122)\cite{K. Gofryk}. Therefore, we cannot rule out the possibility that the observed change in Sommerfeld coefficient is an artifact of the inappropriate subtraction of the phonon contribution. 
$\Delta C/T_c$ was 40~mJ/mol~K$^2$ and 50~mJ/mol~K$^2$ for the as-grown and the annealed crystals, respectively. 
Here, $\Delta$$C$ is the change of electronic specific heat due to superconducting transition.  The present value of $\Delta C/T_c$ is still smaller than the general value of FeSC\cite{Sergey L. Bud' ko}, indicating that the further improvement of the sample quality may be possible.

~~In order to discuss a superconducting gap structure, we examined the magnetic field dependence of $C/T$ at 0~K, $\gamma_r$(H). Although in principle the presence of gap node can be discussed from low $T$ $C(T)$ without magnetic field, it is usually difficult to identify the $T^2$ term which is expected for the superconductor with line node because of the large phonon component\cite{Y. Wang}. Figure 3(a) shows the temperature dependence of $C/T$ at different magnetic fields applied along the $c$-axis. 
In Fig. 3(b) is plotted the field dependences of residual $C/T$, the values at 1.8~K and the extrapolated values at 0~K with $C/T = \gamma_r+\beta T^2$. 
The phonon coefficient $\beta$ also changed from 0.79~mJ/mol~K$^4$ to 0.36~mJ/mol~K$^4$ by annealing, while residual term $\gamma_r$ decreased from 5.5~mJ/mol~K$^2$ to 3.6~mJ/mol~K$^2$ by annealing at $H$=0. This reduction in $\gamma_r$ implies a decrease of residual density of states which originates from impurity or disorder. 
However, the $\gamma_r$ of the annealed crystal is still larger than that of P-Ba122 ($\gamma_r$=1.7~mJ/mol~K$^2$)\cite{J. S. Kim}, indicating a larger residual density of state below $T_c$. 

~~As shown in Fig. 3(b), the field dependence of $\gamma_r$ follows $\gamma_r\propto H^{0.7}$ for the as-grown case, while the post-annealing treatment results in a moderate field dependence, $\gamma_r\propto H$. 
%%%%%%%%%%%%%%%%%%%%%%%%%%%%%%%%%%%%%%%%%%%%%%%%%%%% 
The change from the sub-linear to the $H$-linear dependence of $\gamma_r$ seems to indicate that the superconducting gap structure became a full gap after annealing. However, the theoretical calculation based on the two band model\cite{Y. Bang}, $H$-linear dependence of $\gamma_r$ changes to a sub-linear $H$ dependence when disorders are introduced in a full gap $s_{\pm}$ superconducting state. Therefore, our observation can be understood within a full gap $s_{\pm}$ regime. We note that the similar annealing effect on $\gamma_r$(H) was observed in Co-Ba122\cite{K. Gofryk}. 

~~On the other hand, the penetration depth measurement gave a different result. The observed $\Delta \lambda(T)$ was $T$-linear in the annealed crystal, while $\Delta \lambda(T)$ followed $T^n~(n<2)$ in the as-grown crystal. This change of exponent $n$ can be understood within a framework of nodal superconductor by assuming that the as-grown crystals was in a dirty limit, while the annealed one was in a clean limit. Considering all these results, we speculate that the Fermi surface with a heavier carrier mass has a full gap and dominates the specific heat, while the others with a lighter carrier mass have a nodal gap which governs the penetration depth. Here we note that the $H$-linear dependence of $\gamma_r$ was also observed in P-Ba122 where one Fermi surface has a full gap and the others have nodes\cite{J. S. Kim}.
%%%%%%%%%%%%%%%%crystal structure%%%%%%%%%%%%%%%%%%%%%%%%%%%%%%%%%
%%%%%%%%%%%%%%%%%%%%%%%%
\begin{figure}[h]
\begin{center}
\hspace{2mm}\includegraphics[width=7cm]{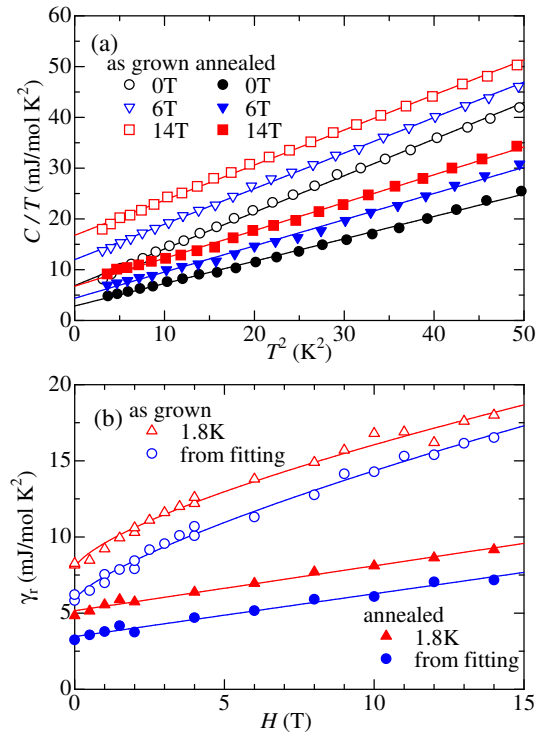}\\
%\vspace{1mm}
%\includegraphics[width=8.5cm]{eps/Fig_01_test2.eps}
\caption{
(Color online)(a) The low temperature specific heat of the as-grown and annealed SrFe$_2$(As$_{0.65}$P$_{0.35}$)$_2$ measured at different magnetic fields applied along $c$-axis. (circles, 0~T; triangles, 6~T; squares, 14~T; open and closed symbols represent the data of the as-grown and the annealed samples, respectively.) (b) The magnetic field dependence of the residual electronic specific heat coefficient, $\gamma_r$. Red triangles and blue circles are correspond to $\gamma_r$ at 1.8~K and $\gamma_r$ from $C/T=\gamma_r+\beta~T^2$, respectively. The solid lines show $\gamma_r\propto H^{0.7}$ for the as-grown and $\gamma_r\propto H$ for the annealed crystal.
}
\label{fig3}
\end{center}
\end{figure}
%%%%%%%%%%%%%%%%%%%%%%%%
%%%%%%%%%%%%%%%%%%%%%%%%%%%%%%%%%%%%%%%%%%%%%%%%%%%%%%%%%%%%%%
\begin{table}[htbp]
\begin{center}
\begin{tabular}{cc|ccccc} 
\hline
%\hline
 \multicolumn{2}{c|}{Compound } &as-grown& annealed   \\%& BaFe$_{2}($As$_{0.71}$P$_{0.29}$)$_2$\cite{BaSrP122_crystal_structure}
\hline
 \multicolumn{2}{c|}{Space group} & $I4/mmm$ & $I4/mmm$    \\ %& $I4/mmm$
 \multicolumn{2}{c|}{$a$ (\AA)} & 3.8983(14) & 3.8963(6)   \\ %& 3.9178(1)
  \multicolumn{2}{c|}{$c$ (\AA)} & 12.064(4) & 12.092(2)   \\ %& 12.7610(7)
  \multicolumn{2}{c|}{$V$  (\AA$^3$)} & 183.33(11) & 183.57(5)   \\ %& 195.9(1)
Sr& &(0, 0, 0) & (0, 0, 0)   \\ %& 2a(0, 0, 0)
Fe& $ $ & (1/2, 0, 1/4) & (1/2, 0, 1/4)   \\ %& 4d(1/2, 0, 1/4)
As/P &  & (0, 0, z) & (0, 0, z) \\ %& 4e(0, 0, z)
 &  & z=0.35931(6) & z=0.35956(5)   \\ %& z=0.3544(1)
$h_\mathrm{Pn}$(\AA)&  & 1.319(1) & 1.325(1)   \\%& 1.332
Bond lengths and angles& &  &  \\ % &  
Sr-As(\AA)& & 3.2372(7) & 3.2364(3)   \\%& 3.3358(7)
Sr-Fe(\AA)& & 3.5910(9) & 3.5964(4)   \\%&
Fe-As/P(\AA)& & 2.3533(4) & 2.3559(3)   \\ %& 2.3689(7)
Fe-Fe(\AA)& & 2.7565(7) & 2.7551(3)  \\%&  
As-Fe-As(deg.)& & 108.301(10)$\times$4 & 108.434(8)$\times$4 \\%& 111.57(5)
               & & 111.84(2)$\times$2 & 111.566(17)$\times$2 \\%&
Number of reflections & & 211 & 219   \\ %& 195.9(1)
(I$>$2.00$\sigma$(I))& & &  \\
Good of fitnesee& &3.315&  6.018 \\
%Goodness of Fittness& &3.315 & 6.018\\
%$R$1& &0.0661&0.0516  \\%& 0.023
 %\multicolumn{2}{c|}{$R_{wp}$ (\%)} & 4.04 & 4.09 & 4.16 \\
 %\multicolumn{2}{c|}{$S$ (\%)} & 2.43 & 2.46 & 2.49  \\
\hline
%\hline
\end{tabular}
\caption{Refined lattice constants, atomic positions, and bond lengths and angles at room temperature for the as-grown and annealed single crystal from the least square refinement of single crystal X-ray diffraction profile. $h_\mathrm{Pn}$ was calculated from $h_\mathrm{Pn}=(z-0.25)\times c$. The reliability are $R_1~(I>2.00\sigma(I))=6.61\%, 5.16\%$ and $wR_2~(I>2.00\sigma(I))=10.17\%, 8.15\%$ for the as-grown and annealed crystal, respectively. Number of reflections is number of diffraction peaks used for analysis.}
\label{crystalstructure}
\end{center}
\end{table}
%%%%%%%%%%%%%%%%%%%%%%%%%%%%%%%%%%%%%%%%%%%%%%%

~~Finally to understand the origin of electronic change, we examine the crystal structure. Table 1 shows the annealing effect on the crystal structure. In the analysis, we adjusted the coordinate $z$ of As/P simultaneously. The structure analysis clearly demonstrates that the annealed crystal has the shorter $a$-axis and longer $c$-axis than the as-grown crystal. In addition, the $z$ position of As/P also increases, leading to a higher pnictogen height $h_\mathrm{Pn}$ and a smaller As-Fe-As bond angle.
In FeSC, it has been pointed out that $T_c$ is correlated with the pnictogen height and/or the As-Fe-As bond angle\cite{C.-H. Lee, Y. Mizuguchi}. Considering that $T_c$ sharply changes around $h_\mathrm{Pn}\sim1.33~\rm\AA$\cite{Y. Mizuguchi}, a tiny extension of $h_\mathrm{Pn}$ in SrFe$_2$(As$_{0.65}$P$_{0.35}$)$_2$ ($\sim1.32~\rm\AA$) can result in an enhancement of $T_c$. 
Another important effect of post-annealing is to decrease disorders within crystals. The related phenomena were that annealing decreased $\gamma_r$ and slightly increased RRR. Considering that the phonon term in specific heat also changed by annealing, we suppose that the lattice distortion or inhomogeneity of As/P acted as disorders. These disorders could cause the mass enhancement of carriers and the larger Sommerfeld coefficient in the as-grown crystal, while it can be suppressed by annealing. The reduction of disorders made the system cleaner and modified the gap feature. The observed $T_c$-enhancement also may be partially caused by the reduction of carrier scattering.
%%\subsection{Summary}

~~In summary, we studied the annealing effect on P-Sr122 and performed magnetic susceptibility, electric resistivity, specific heat measurements and a precise X-ray diffraction analysis. We found that $T_c$ is enhanced up to 33~K that is higher than the value of optimal P-Ba122. We found that there are two effects of post-annealing. One is the elongation of the $c$-axis and the pnictogen height in the average crystal structure. This causes the enhancement of $T_c$ in the annealed crystals. The other effect is to reduce disorders within crystals, which is clearly observed in specific heat. This suggests that the system changes from a dirty to clean superconductor with the $s_{\pm}$ full gap on the Fermi surface with a heavier carrier mass.
%\vspace{0.2in}\\*

~~We thank Y. Wakabayashi for his technical support in the X-ray diffraction analysis. X-ray diffraction experiment has been carried out under approval of the Photon Factory Program Advisory Committee (personal Nos. 2009S-008 and 2012S-005). The present work was supported by Scientific Research S (21224008), by JSPS, FIRST program, and by JST, CREST, TRIP, IRON-SEA.

%%%%%%%%%%%%reference%%%%%

\end{document}